\documentclass[aps,prl,twocolumn,showpacs]{revtex4}

\usepackage{graphicx}

\begin{document}

\title{Tuning the Mott transition in a Bose-Einstein condensate
by multi-photon absorption}

\author{C.E.~Creffield and T.S.~Monteiro}
\affiliation{Department of Physics and Astronomy, University College London,
Gower Street, London WC1E 6BT, United Kingdom}

\date{\today}

\begin{abstract}
We study the time-dependent dynamics of a Bose-Einstein condensate
trapped in an optical lattice. Modeling the system as a Bose-Hubbard
model, we show how applying a periodic driving field
can induce coherent destruction of tunneling.
In the low-frequency regime, we obtain the novel result that the 
destruction of tunneling displays extremely sharp peaks
when the driving frequency is resonant with the depth of the 
trapping potential (``multi-photon resonances''), which allows
the quantum phase transition between the Mott insulator 
and the superfluid state to be controlled with high precision. 
We further show how the waveform of the field can be chosen to
maximize this effect.
\end{abstract}


\maketitle

Recent spectacular progress in trapping cold atomic gases \cite{review}
has provided a new arena for studying quantum many-body physics.
In particular, ultracold bosons held in optical potentials provide an
almost ideal realization of the Bose-Hubbard (BH) model \cite{jaksch},
in which the model parameters can be controlled to high precision.
As well as their purely theoretical interest, these systems
attract attention because of their possible
application to quantum information processing \cite{jacksh_comp}.

The BH model is described by the Hamiltonian
\begin{equation}
H_{BH} = -J \sum_{\langle i, j \rangle} \left[ a_i^{\dagger} a_j + H.c. \right]
+ \frac{U}{2} \sum_i n_i \left( n_i - 1 \right),
\label{bh-ham}
\end{equation}
where $a_i / a_i^{\dagger}$ are the standard annihilation/creation
operators for a boson on site $i$,
$n_i = a_i^{\dagger} a_i$ is the number operator,
$J$ is the tunneling amplitude between neighboring sites, and $U$ is the
repulsion between a pair of bosons occupying the same site.
Its physics is governed by the competition
between the kinetic energy and the Hubbard interaction,
and thus by the ratio $U/J$.
When $U /J \ll 1$ the tunneling dominates, and the ground
state of the system is a superfluid.
As $U/J$ is increased the system passes through
a quantum phase transition, and evolves into a Mott-insulator (MI)
state in which the bosons localize on the lattice sites.

This phase transition was observed experimentally in Ref.\cite{greiner}
by varying the depth of the optical potential.
In this Letter we propose an alternative method:
applying an additional oscillatory potential induces
{\em coherent destruction of tunneling} (CDT), and thus 
suppresses the effect of  $J$.
CDT is a quantum interference effect, discovered in the pioneering
work of Ref.\cite{hanggi}, in which the period for tunneling between
states diverges as their associated quasienergies \cite{hanggi_review}
approach degeneracy.

Here we show how CDT can be used to 
control the dynamics of a boson condensate,
by means of a novel resonance effect between $U$ and the frequency
of the driving field.
We consider a one-dimensional BH model, driven by a time-periodic
potential which varies linearly with site number. The Hamiltonian is
given by:
\begin{equation}
H(t) = H_{BH} +	 K f(t) \sum_j^N \ j \	n_j ,
\label{drive}
\end{equation}
where $K$ is the amplitude of the driving field,
and $f(t)$ is a $T$-periodic function of
unit amplitude that describes its waveform.
Such time-periodic linear potentials -- generated
by an accelerated lattice for example --
have already been used in cold-atom
experiments \cite{Creff05}.
A similar form of driving-potential was also recently
investigated theoretically  \cite{eckardt} in the high-frequency regime 
($\omega > U$), and 
was found to suppress the transition to the superfluid regime.

\begin{center}
\begin{figure}
\includegraphics[width=0.45\textwidth,clip=true]{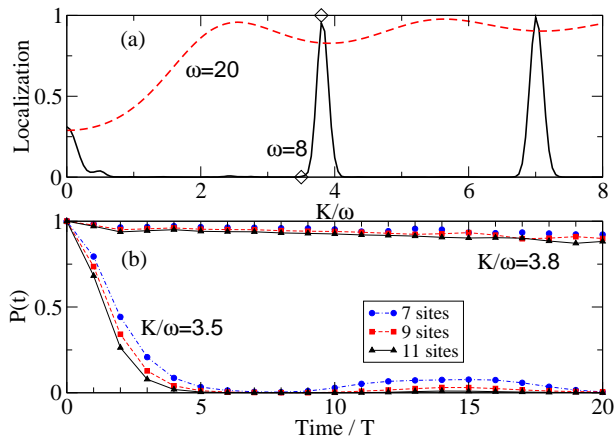}
\caption{(a) The minimum overlap with the MI-state,
or {\em localization}, reached in
a 7-site system with $U=8$, during 10 periods of driving. For $\omega=20$
(dashed-line)
the localization peaks at $K/\omega=2.4, 5.5$ -- the zeros
of ${\cal J}_0(K/\omega)$. When $\omega$ is reduced
to $\omega = 8$ (the first photon resonance, solid-line),
the peaks become extremely narrow and are centered
on the zeros of ${\cal J}_1(K/\omega)$. The diamonds mark the points
$K/\omega=3.5$ and 3.8 (see below).
(b) Time evolution of the $\omega=8$ case for three system sizes,
7, 9 and 11 sites.
For $K/\omega=3.5$, away from the resonance, the overlap with the initial
state, $P(t)$, rapidly drops to zero. For $K/\omega=3.8$, at
the peak of the resonance, the decay is much slower, indicating that
the driving field preserves the MI-state.}
\label{sine}
\end{figure}
\end{center}

Here, for the first time, the multi-photon (low-frequency) regime
is investigated. An unexpected new finding is that CDT is now
modulated by a set of extremely sharp `resonances' (the contrast between the
high-frequency behavior and the multi-photon regime investigated here
is illustrated in Fig.\ref{sine}a). This means that the Mott-transition
can be induced by minute changes in experimental parameters.

Henceforth we put $\hbar=1$ and measure all energies in units of $J$,
and set the number of bosons equal to the number of lattice sites $N$.
Although the dimension of the Hilbert space increases
exponentially with $N$, in a Fock basis
$H$ is extremely sparse, with {\em at most} $(2N-1$) non-zero entries
per row. Thus despite the rapid increase in the dimension of
the Hilbert space, this sparsity allows us to treat relatively
large systems of up to eleven sites, and so assess if the effects
we observe survive in the thermodynamic limit.

Our numerical investigation consists of initializing the system
in the `ideal' MI state,
$| \Psi_{\mbox{\small MI}} \rangle = \prod a^{\dagger}_j | 0 \rangle$,
and evolving the many-particle Schr\"odinger
equation over time (typically ten periods of the
driving field) using a Runge-Kutta method.
To study the system's time-evolution quantitatively,
we measure the overlap of the wavefunction with
the initial state $P(t) = | \langle \Psi_{\mbox{\small MI}} |
\Psi(t) \rangle|^2$. For convenience we term the minimum value of $P(t)$
attained during the time-evolution to be the {\em localization}.
When CDT occurs,
the system will remain frozen in the MI-state, and consequently
the localization will be close to one. Conversely, if the bosons
are able to tunnel freely from site to site, the value
of the localization will be reduced.

We begin by considering the case of sinusoidal driving, $f(t)=\sin \omega t$.
The MI transition is quite soft in 1D, starting
at $U \simeq 4$ and developing fully for
$U > 20$. Throughout this work we use an intermediate value of $U=8$.
Fig.\ref{sine}a shows how the localization in a
7-site system varies as the amplitude of the driving field is increased,
while its frequency is held constant at $\omega=20$.
For $K=0$ the localization has a value of $\sim 0.3$, demonstrating that
in the absence of a driving field this value of $U$ is indeed
insufficient to maintain the MI-state.
Applying the driving field
causes the localization to steadily rise from this value as $K$ is increased
from zero, indicating that the effective tunneling between lattice sites
is increasingly suppressed,
until it peaks at a value close to one at $K/\omega=2.4$.
As $K$ is increased further, the localization goes through a 
shallow local minimum, before again peaking at $K/\omega=5.5$. 
It was observed \cite{eckardt} that these values
of $K/\omega$ are close to the first two zeros of ${\cal J}_0$,
the zeroth Bessel function.

Reducing the driving frequency to a lower value, $\omega=8$,
produces a radically different behavior --
the value of the localization
rapidly {\em drops} as $K/\omega$ is increased from zero,
indicating that the field {\em destroys} the MI-state.
As $K/\omega$ is increased further the value of the localization
remains extremely low except at a series of very sharp peaks.
Fig.\ref{sine}b emphasizes the narrowness of these peaks by showing
the time-evolution of the system for two values of $K$.
For the first, $K=3.5 \omega$, $P(t)$ rapidly
falls from its initial value, reaching a level near zero within
five driving periods. There is a small dependence on the system
size, with the decay occurring more quickly as $N$ is increased.
At the localization-peak,
$K = 3.8 \omega$, $P(t)$ decays far more slowly with time,
so that after twenty periods of driving it only falls
to a value of $\sim 0.9$, and only minor dependence on $N$ is evident.
Thus for this value of $\omega$, altering the amplitude
of the field by just $10 \%$ produces enormous differences in the
localization.

Although the Hamiltonian (\ref{drive}) is explicitly
time-dependent, the fact that it is periodic allows us
to use the Floquet theorem to write solutions of the Schr\"odinger equation
as $\psi(t) = \exp\left[-i \epsilon_j t \right] \phi_j(t)$,
where $\epsilon_j$ is the quasienergy, and $\phi_j(t)$ is a
$T$-periodic function called the Floquet state \cite{hanggi_review}.
As the quasienergies are only defined
up to an arbitrary multiple of $\omega$, the quasienergy
spectrum possesses a Brillouin zone structure, in precise analogy
to the quasimomentum in spatially periodic crystals.
For the Floquet analysis, we work in an
{\em extended} Hilbert space of $T$-periodic functions \cite{sambe}.
In this approach, the Floquet states and quasienergies satisfy
\begin{equation}
{\cal H}(t) | \phi_j(t) \rangle
= \epsilon_j | \phi_j(t) \rangle \ ,
\label{eval}
\end{equation}
where ${\cal H}(t) = H(t) - i \hbar {\partial}/{\partial t}$.
Working in this extended Hilbert space thus reduces the task
of calculating the time-dependent, driven dynamics of the system
to a time-{\em independent} eigenvalue problem.

To study the behavior of the quasienergies,
we make use of a perturbative scheme developed in \cite{holthaus}
to treat non-interacting systems, and later generalized in \cite{creff}
to include interactions. Our procedure is to first find the eigensystem of the
operator ${\cal H}_0(t) = H_0(t) - i \hbar {\partial}/{\partial t}$,
where $H_0$ contains terms diagonal
in a Fock basis (i.e. the driving term
and the Hubbard interaction). We are then able to 
use standard Rayleigh-Schr\"odinger perturbation theory
to evaluate the corrections to this result, using
the remaining terms of $H_{BH}$ as the perturbation.

For the two-site system, a natural basis is given by the Fock
states $\{ |1,1 \rangle, \ |2,0 \rangle,\ |0,2 \rangle\}$,
where $|n,m \rangle$ denotes the state with $n$ bosons on
the first site and $m$ on the second. Finding the eigensystem
of ${\cal H}_0$ then amounts to solving three first-order differential
equations, yielding the result:
\begin{eqnarray}
|\phi_+(t)\rangle = \left(0, \exp \left[-i (U - \epsilon_+) t
+ i \frac{K}{\omega} \cos \omega t \right], 0 \right) \nonumber \\
|\phi_0(t)\rangle = \left(0, 0, \exp \left[-i (U - \epsilon_0) t
- i \frac{K}{\omega} \cos \omega t \right] \right) \nonumber \\
|\phi_-(t)\rangle = \left(\exp \left[i \epsilon_- t \right], 0, 0 \right) .
\qquad \qquad 
\end{eqnarray}
Imposing the $T$-periodic boundary condition on these states
requires setting $\epsilon_- = 0$ and
$(U -\epsilon_{+ / 0}) = m \omega$, where $m$ is an integer.
Thus in general it is not possible to include
the full Hubbard-interaction term within $H_0$,
depending on its commensurability with $\omega$.
To deal with this it is necessary to
decompose $U$ into a form which duplicates the
Brillouin zone structure of the quasienergies
\begin{equation}
U = n \omega + u, \quad n=0,1,2\dots
\label{brill}
\end{equation}
where $u$ is the ``reduced interaction'', $|u| \leq \omega/2$.
This decomposition reveals that
only the reduced interaction need be included in the
perturbation, while the remainder of $U$ (an integer multiple
of $\omega$) can be retained in $H_0$.

\begin{center}
\begin{figure}
\includegraphics[width=0.45\textwidth,clip=true]{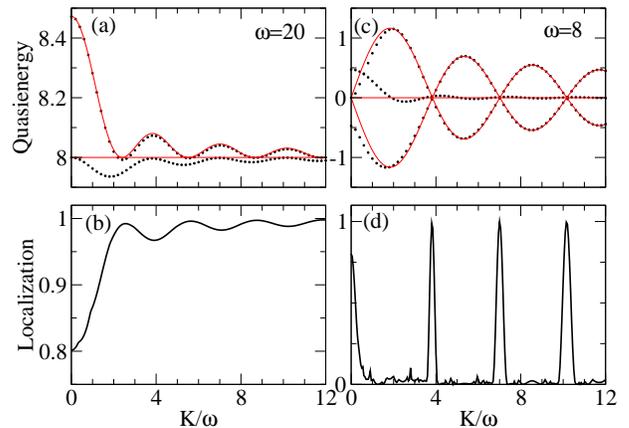}
\caption{(a) Quasienergy spectrum of a 2-site system,
with $U=8$ and $\omega=20$. Only two of the three
quasienergies are plotted (the remaining one oscillates
weakly about zero), which make a series of close
approaches to each other as $K$ is increased.
Red-lines denote the perturbative solutions, which agree
well with the exact results (black circles).
(b) At the points of close approach, the tunneling
is suppressed and the localization peaks.
For all field strengths the tunneling is
suppressed with respect to the undriven system, and the
localization is thus enhanced.
(c) At lower frequencies the behavior
of the quasienergies changes dramatically. At $\omega=8$ the system
is at the first photon resonance ($U = \omega$), and
the behavior of the quasienergies is described extremely
well by the perturbative solutions $\epsilon = 0, \pm2 {\cal J}_1(K/\omega)$.
(d) As before, the localization is peaked at
the points of quasienergy crossing, but in contrast
to (b) the peaks are extremely sharp.}
\label{twosite}
\end{figure}
\end{center}

To first-order it is easily shown that the
three quasienergies are given by
\begin{equation}
\epsilon_0 = u \quad \mbox{and} \quad
\epsilon_{\pm} = \left(u \pm \sqrt{u^2 + 16 J_{\mbox{\small{eff}}}^2 }\right)/2 \ ,
\label{quasi}
\end{equation}
where the intersite tunneling has been reduced to an
effective value $J_{\mbox{\small{eff}}} = J \ {\cal J}_n(K/\omega)$,
and $n$ and $u$ are defined in Eq.\ref{brill}.
Thus in the high-frequency limit ($\omega \gg U$), when $n=0$
and $u = U$, it is clear that
the quasienergies $\epsilon_0$ and $\epsilon_+$ are degenerate
when ${\cal J}_0(K/\omega)=0$.
In Fig.\ref{twosite}a we show the excellent agreement between the
perturbative result and the exact quasienergies for a driving
field of frequency $\omega=20$. Below, in Fig.\ref{twosite}b,
we plot the corresponding value of the localization, and it can be
clearly seen that the peaks in this quantity are indeed centered on the points
of closest approach of the quasienergies.
It may be seen from Eq.\ref{quasi} that for large
values of $u$, the quasienergy separation 
$(\epsilon_+ - \epsilon_0) \simeq 4J_{\mbox{\small{eff}}}^2/u$. The effect 
of $u$ is thus to reduce the amplitude of oscillations in this quantity, 
and so to smear out the avoided crossings of the quasienergies.
As a result, the peaks in the localization are rather broad and overlap
each other, and thus
the localization cannot reach particularly low values.

Eq.\ref{brill} reveals the particular importance
of {\em photon resonances}, when $U$ is an integer
multiple of the frequency of the driving field, $U = n \omega$.
When this condition is satisfied the reduced interaction is
zero, and the crossings between the quasienergies are
well-defined. This is the origin of the extremely sharp peaks
in localization seen in Fig.\ref{sine}a for $\omega = U =8$.
Away from these peaks, the photon-absorption compensates
for the energy cost of doubly-occupying a lattice site,
in analogy to the photon-assisted tunneling studied in Ref.\cite{pat},
thereby producing low values of localization.
In Fig.\ref{twosite}c we plot the quasienergies for the
first photon resonance ($n=1$) for the two-site system.
For weak fields ($K/\omega < 2$)
small deviations of the exact quasienergies
from the perturbative result are visible, but for higher
field strengths the agreement is again excellent. In Fig.\ref{twosite}d we
plot the localization produced in this system, and we can note
that, as seen previously in the 7-site system, the localization takes
extremely low values except at a set of very narrow peaks. These
peaks are precisely aligned with the quasienergy crossings at the
zeros of ${\cal J}_1(K/\omega)$.

\begin{center}
\begin{figure}
\includegraphics[width=0.45\textwidth,clip=true]{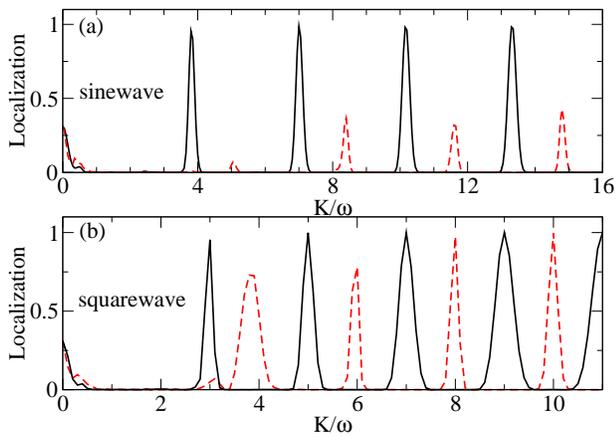}
\caption{Localization produced in
a 7-site system ($U=8$) for two forms of periodic driving:
(a) sinusoidal, (b) squarewave.
Both waveforms produce excellent localization at the first
photon resonance, $\omega=U$, shown by solid-black lines.
At the second photon resonance ($2\omega=U$), shown by red-dashed lines,
the localization
produced by the sinusoidal driving is considerably smaller,
but the squarewave still produces high peaks.}
\label{compare}
\end{figure}
\end{center}

As we reduce $\omega$ still further, we can expect to encounter
a sequence of higher resonances with similar behavior.
In Fig.\ref{compare}a we compare the values of localization produced
in a 7-site system for the $n=1$ and $n=2$ resonances.
The second photon resonance, however, produces a
worse result than for $n=1$. Although sharp
peaks are still present in the localization, and in agreement with the
perturbation theory they are indeed
centered on the zeros of ${\cal J}_2(K/\omega)$,
the maximum value of localization produced is considerably lower.

This poor localization occurs because not all
of the quasienergy crossings in the Floquet spectrum
of a many-site system will occur at precisely the same
value of $K/\omega$; instead the various crossings occur
over an interval. Thus although at the peaks in
the localization many pairs of Floquet states will be degenerate
(and tunneling between them will be suppressed), other
state-pairs will only be {\em approximately degenerate} and will permit
a small, but non-zero, degree of tunneling to occur.
The major factors influencing this effect 
arise from higher-order terms in the expansion
of the single-period time-evolution operator $U(T,0)$, which
manifest as multi-particle tunneling and tunneling
beyond nearest neighbors.
CDT is a quantum interference effect, which occurs when
the dynamical phase acquired by a particle in a period of driving
produces destructive interference, thereby suppressing
the particle's dynamics. If, however, a `clump' of $n_1$ bosons tunnels
between sites, the dynamical phase will be $n_1$ times larger
than that for a single boson.
Similarly, if a boson tunnels between two sites separated
by $n_2 > 1$ lattice spacings, the dynamical phase will be $n_2$ times larger.
For sinusoidal driving, the single-particle tunneling is
suppressed when ${\cal J}_n (K/\omega)=0$; for these higher-order
processes also to be suppressed we therefore also require
${\cal J}_n (n_1 n_2 K/\omega) = 0$
for integers $n_1,n_2=1,2\dots N$.

Clearly this condition cannot be satisfied for sinusoidal driving,
as the zeros of ${\cal J}_n (x)$ are not equally spaced.
Thus to observe good localization properties at high photon-resonances
we need to to construct
a driving field $f(t)$ such that the crossings in its
Floquet spectrum are periodically spaced. This problem was
confronted in a different context in Ref.\cite{dignam},
where it was shown that such a field
must be {\em discontinuous} at changes of sign.
Possibly the simplest field of this type, and the most convenient
for experiment, is a squarewave field.

\begin{center}
\begin{figure}
\includegraphics[width=0.45\textwidth,clip=true]{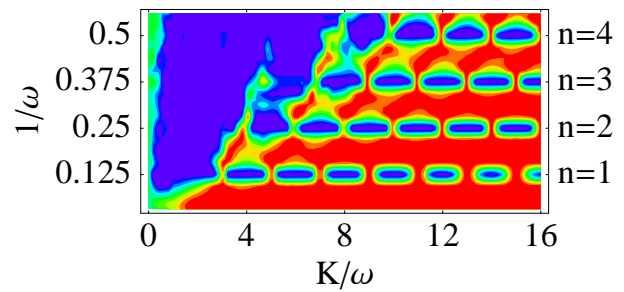}
\caption{The localization produced in a
5-site system with $U=8$ as a function of the
frequency $\omega$ of the squarewave driving field and its
amplitude $K$. When $n \omega = U$
the localization is almost
zero (the horizontal blue bands) except at sharply defined peaks (red).
Between the bands localization is good (red).
The $n=1,2,3$ and $4$ resonances are marked on the right.
The upper-left triangle (blue) displays poor localization and
little structure, corresponding to the non-perturbative regime.}
\label{photon}
\end{figure}
\end{center}

In Fig.\ref{compare}b we show the localization obtained in a
$7$-site system driven by a squarewave.
Using the same perturbative approach as before, it may be shown
that the quasienergy degeneracies
occur for $(K/\omega) = 2m + 1 \mbox{ \ or \ } 2 m$, depending
on whether the order of the resonance, $n$, is odd or even.
Unlike the case of sinusoidal driving,
the $n=2$ resonance displays good localization,
comparable to that obtained for $n=1$.
A contour plot showing the localization as a function of
both $K/\omega$ and $\omega^{-1}$ is presented in Fig.\ref{photon}.
The prominent horizontal bands correspond to the photon
resonances ($\omega^{-1} = n/U$), which are punctuated
by a series of narrow peaks at which the
localization is preserved. This plot also clearly shows
the division between the fairly featureless, poorly-localized,
``weak-driving'' regime
to the upper left, and the ``strong-driving'' regime which shows
the resonance features. For the latter,
the dynamics of the system are dominated by
the combined effect of the driving field and the Hubbard interaction,
and thus is well-described by our form of perturbation theory.
Fig.\ref{photon} allows us to locate the boundary between
the two regimes quite accurately as $K/\omega \simeq (2 U)/\omega$.

In summary, we have investigated the dynamics of the BH
model under a periodic driving field. For high frequencies
\cite{eckardt} the field can be used to inhibit tunneling by means of CDT
and thus stabilize the MI-state. Lowering the frequency, however,
reveals the existence of resonance effects which can be used
to selectively destroy {\em or} preserve the MI-state.
Lower driving frequencies have the added advantages in experiment
that they heat the condensate less, and will not drive
transitions to higher Bloch bands thereby
invalidating the single band model.
The extremely narrow width of the resonance features indicates
that it should be possible to control the Mott transition very precisely
in this manner.


\begin{thebibliography}{99}
\bibitem{review}
{O.~Morsch and M.~Oberthaler,
Rev. Mod. Phys. {\bf 78}, 179 (2006).}

\bibitem{jaksch}
{D.~Jaksch  {\it et al.}
Phys. Rev. Lett. {\bf 81}, 3108 (1998).}

\bibitem{jacksh_comp}
{D.~Jaksch {\it et al.}, 
Phys. Rev. Lett. {\bf 82}, 1975 (1999).}

\bibitem{greiner}
{M.~Greiner {\it et al.},
Nature {\bf 415}, 39 (2002).}

\bibitem{hanggi}
{F.~Grossmann, T.~Dittrich, P.~Jung, and P.~H\"anggi, Phys. Rev. Lett.
{\bf 67}, 516 (1991).}

\bibitem{hanggi_review}
{M.~Grifoni and P.~H\"anggi, Phys. Rep. {\bf 304}, 229 (1998).}

\bibitem{Creff05}
{G.~Hur, C.E.~Creffield, P.H.~Jones, and T.S.~Monteiro,
Phys. Rev. A {\bf 72}, 013403 (2005).}

\bibitem{eckardt}
{A.~Eckardt, C.~Weiss, and M.~Holthaus,
Phys. Rev. Lett. {\bf 95}, 260404 (2005).}

\bibitem{sambe}
{H.~Sambe, Phys. Rev. A {\bf 7}, 2203 (1973).}

\bibitem{holthaus}
{M.~Holthaus, Z. Phys. B {\bf 89}, 251 (1992).}

\bibitem{creff}
{C.E.~Creffield and G.~Platero, Phys. Rev. B {\bf 65}, 113304, (2002);
C.E.~Creffield and G.~Platero, Phys. Rev. B {\bf 66}, 235303 (2002).}

\bibitem{pat}
{A.~Eckardt, T.~Jinasundera, C.~Weiss, and M.~Holthaus,
Phys. Rev. Lett. {\bf 95}, 200401 (2005).}

\bibitem{dignam}
{M.M.~Dignam and C.~M.~de~Sterke, Phys. Rev. Lett. {\bf 88},
046806 (2002).}

\end{thebibliography}
\end{document}